\documentclass[a4paper]{VisionStyle}
\usepackage{epsfig}

\begin{document}

\title{Status of the EPIC/MOS calibration}

\author{P.\,Ferrando\inst{1} \and A.F.\,Abbey\inst{2} \and 
        B.Altieri\inst{3} \and M.\,Arnaud\inst{1} \and 
        P.\,Bennie\inst{2} \and M.\,Dadina\inst{4}  \and 
	M.\,Denby\inst{2} \and S.Ghizzardi\inst{5} \and 
	R.G.\,Griffiths\inst{2} \and N.\,La Palombara\inst{5} \and
	A.\,de~Luca\inst{5} \and D.\,Lumb\inst{6} \and 
	S.\,Molendi\inst{5} \and D.\,Neumann\inst{1} \and 
	J.L.\,Sauvageot\inst{1} \and R.D.\,Saxton\inst{3,7} \and 
	S.\,Sembay\inst{2} \and A.\,Tiengo\inst{3,5} \and 
	M.J.L.\,Turner\inst{2} } 

\institute{
  Service d'Astrophysique, CEA/Saclay, 91191 Gif-sur-Yvette Cedex, 
  France
\and
Dept.\ of Physics and Astronomy, Leicester University, Leicester LE1 
7RH, U.K.
\and 
  XMM-SOC Satellite tracking station, Villafranca del Castillo, 28080 
  Madrid, Spain
  \and 
Istituto Te.S.R.E, CNR, via Gobetti 101, 40129 Bologna, Italy
\and
IASF-Sezione di Milano, 20133 Milano, Italy
\and
ESA Astrophysics Missions Division, Research and Scientific Support 
Department, ESTEC, Postbus 299, NL-2200 AG Noordwijk, The Netherlands
\and
XMM-SSC, Dept.\ of Physics and Astronomy, Leicester University, Leicester LE1 
7RH, U.K.}

\maketitle 

\begin{abstract}

The XMM-Newton observatory has the largest collecting area flown so 
far for an X-ray imaging system, resulting in a very high sensitivity 
over a broad spectral range. In order to exploit fully these 
performances, a very accurate calibration of the XMM-Newton 
instruments is required, and has led to an extensive ground and flight 
calibration program. We report here on the current
status of the EPIC/MOS cameras calibrations, highlighting areas for 
which a reasonably good accuracy has been achieved, and noting points 
where further work is needed.

\end{abstract}

\section{Introduction}
\label{pferrando-WA2_sec:intro}

In order to achieve the highest collecting power ever deployed in X-ray 
astronomy, the XMM-Newton observatory carries three telescopes with 
identical mirrors 
(\cite{pferrando-WA2:Jansen2001}). There is one CCD camera at the focus of 
each mirror, the ensemble of these three cameras making up the European Photon
Imaging Camera instrument (EPIC, \cite{pferrando-WA2:Turner2001}). Two of the 
telescopes have X-ray gratings mounted behind the mirrors, and for them 
the original flux is roughly equally shared between the EPIC focal plane and 
the Reflection Grating Spectrometer (RGS, \cite{pferrando-WA2:denHerder2001}); 
the two corresponding EPIC cameras are made of MOS CCDs 
(\cite{pferrando-WA2:Short1998}), and are the ones we will be discussing in 
this paper. The third telescope has an unobstructed beam, and is equipped 
with PN CCDs (\cite{pferrando-WA2:Strueder2001}). The calibration status 
of this EPIC/PN camera is discussed by \cite*{pferrando-WA2:Briel2002} in 
these proceedings.

The overall response of the MOS cameras, which is the final product 
of interest for the observer when analysing his data, results from 
the combination of the calibration of all the individual hardware parts 
which are encountered by the X-rays between the celestial source and the focal 
plane detectors. These hardware pieces, and the associated 
calibration items are~:
\begin{itemize}
  \item the mirror, characterized by its on-axis effective 
  area, vignetting, and point spread function,
  \item the grating stack of the RGS (not present in the PN camera), 
  characterized by its transmission function, and its azimuthal 
  modulation,
  \item the filter inside the EPIC camera, which can be either thin, 
  medium or thick, characterized by its transmission and its 
  homogeneity,
  \item the focal plane detectors, characterized by the quantum efficiency, 
  charge transfer inefficiency, homogeneity, redistribution matrix, energy 
  scale of the CCDs, as well as their metrology.
\end{itemize}
In addition to measuring the above characteristics, the calibration tasks 
also include the two problems of i) the background determination, 
essential for the study of extended sources, and ii) 
pile-up treatment, essential for the study of strong point sources.

Prior to launch, a large number of subsystem calibrations have been determined 
by ground measurements. The most outstanding are the mirrors characterization 
in UV and X-rays at the CSL and Panter facilities respectively, as 
well as the extensive CCDs measurements performed in the ORSAY/LURE 
synchrotron beams (e.g.\ \cite{pferrando-WA2:Pigot1999}). If these ground results have 
given a solid basis for building the calibration data base, they had obviously 
first to be checked and refined with flight data. Furthermore, given the 
few percent accuracy goal set for the calibrations, a number of 
calibration aspects were not entirely, or not satisfactorily, covered by 
the ground measurements; this is the case for example of the CCD quantum 
efficiency at low energy. Some others, like the metrology or 
background aspects, and obviously the monitoring of the CCD Charge 
Transfer Inefficiency degradation, can be studied with flight data only. 
This is the reason for the intensive flight calibration program which has 
followed the commissioning phase, and for the routine calibration observations 
which are scheduled (see list in \cite{pferrando-WA2:Ehle_Altieri2001}). 
A first report on these in-flight calibration activities was given by 
\cite*{pferrando-WA2:Lumb2000}. Although the exercise is a rather drastic 
simplification of the complex calibration analysis, 
Table~\ref{pferrando-WA2:tab1} gives a first cut estimate of the 
main contributor, ground or flight data, to each of the calibration items 
mentioned above.

\begin{table}[ht]
  \caption{First order calibration matrix, indicating the main 
  source of data used for filling the Current Calibration Files.}
  \label{pferrando-WA2:tab1}
  \begin{center}
    \leavevmode
    \footnotesize
    \begin{tabular}[h]{llcc}
      \hline \\[-5pt]
                 & Item           & Ground  &  Flight \\[+5pt]
      \hline \\[-5pt]
 Telescope  & On axis area        &   X    &          \\
            & Point Spread Function &      &    X     \\
	    & Vignetting          &        &    X     \\
 Gratings   & Transmission        &   X    &          \\
            & Azimuthal Modulation &   X   &          \\
 Filters    & Transmission        &   X    &         \\
	    & Homogeneity         &   X    &         \\	 
 CCDs       & Quantum efficiency  &   X    &    X     \\
            & Charge Transf.\ Ineff.\ &    &    X     \\
	    & Redistribution Matrix &  X   &    X     \\
	    & Homogeneity         &   X    &          \\
	    & Background          &        &    X     \\
	    & Metrology, astrometry &      &    X     \\
	    & Timing              &        &    X     \\
      \hline \\
      \end{tabular}
  \end{center}
\end{table}
In this report, and after a brief section on the basic 
operating modes of the MOS cameras, we will mainly describe the 
status of the items listed under the ``flight data" column in 
Table~\ref{pferrando-WA2:tab1}.

\begin{figure}[ht]
  \begin{center}
    \epsfig{file=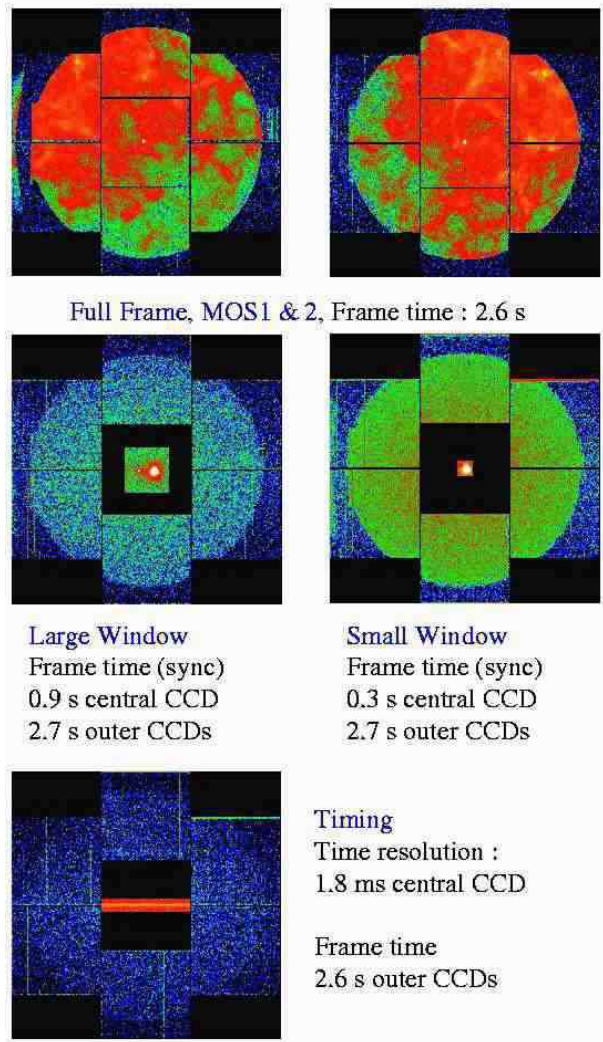, width=8cm}
  \end{center}
\caption{The four basic modes of the MOS cameras.}  
\label{pferrando-WA2:fig-modes}
\end{figure}

\section{The EPIC/MOS cameras and modes}
\label{pferrando-WA2_sec:cameras}

The two EPIC/MOS cameras are made of 7 CCDs each, passively cooled to 
an operating temperature of $-100\degr~C$. Each CCD is operated in a 
frame transfer mode, and has a $600 \times 600$ pixels imaging 
section, the pixel size being of 40~$\mu$m, corresponding to 1.1\arcsec
on the sky. The CCDs are arranged with a central CCD and 6 peripheral 
CCDs, as can be seen in Figure~\ref{pferrando-WA2:fig-modes}. There are 
three basic imaging modes for the operation of the camera, and one 
timing mode. Images taken in these modes are displayed in 
Fig.\ref{pferrando-WA2:fig-modes}; the two top images correspond to 
MOS1 and MOS2 for the same field, and illustrate the fact 
the two MOS cameras are oriented at $90 \degr$ from each other. For 
the timing mode, the short axis on the central chip image corresponds to 
a true spatial coordinate axis, while the long axis coordinate is 
here a measurement of time, and therefore carries no spatial information. 
Note that the peripheral CCDs are always operated in 
Full Frame, whatever the camera mode.

\section{Imaging characteristics}
\label{pferrando-WA2_sec:charact}

\subsection{Metrology and absolute astrometry}
\label{pferrando-WA2_sec:metrology}
Contrary to the PN camera which has a monolithic CCD, the seven CCDs 
of the MOS camera are mounted individually, and it was necessary to 
measure their relative positions, in X, Y, and rotation angle using 
celestial sources. The method for refining the MOS metrology 
involves iteratively improving boresight and CCD offset values in 
the Current Calibration Files (CCFs) for both cameras, using known 
star fields.
Starting with the most recent iteration of the CCFs calibrated event 
lists were produced using the MOS processing chain in the SAS, which 
were in turn used for generating cleaned, band limited, images suitable 
for source searching. The SAS source detection chain was then used to 
generate source lists. Using the maximum likelihood this source list 
was then used in conjunction with the USNO catalogue positions to 
i) remove gross offsets in the central CCDs via shifts in the 3 
boresight Euler angles, and ii) optimise the outer CCD displacements
by minimising the rms offset error between the detected positions 
and catalogue positions for each CCD.

This processing was performed on the Orion Molecular Cloud field, the 
\object{NGC 2516} open cluster, and the Lockman Hole field. 
This has allowed the source detection error to be reduced to approx 
0.4\arcsec and the rms offset errors between detections and catalogue 
positions to approx 0.5\arcsec for MOS1 and 0.55\arcsec for MOS2. 
This accuracy, which is half the pixel size, is probably at the limit 
of this single field method and further improvements in the MOS metrology 
will have to come through statistical studies of large numbers of 
detections over the lifetime of XMM-Newton. This work has now been 
fully reflected in the CCFs.

It must be clear that these numbers characterize the internal focal 
plane metrology, and not the absolute pointing accuracy reached in the 
absence of any positional reference in the X-ray image. This last question 
has been addressed by \cite*{pferrando-WA2:Tedds_Watson2002} in these 
proceedings, who performed a statistical study on a very large 
number of fields; they have found that the absolute astrometric error 
of the XMM-Newton coordinate frame has a distribution with a FWHM of 
$\sim$~3--4\arcsec per axis (R.A.\ and Dec), and 
0.3\degr in roll angle.

\subsection{Point Spread Function}
\label{pferrando-WA2_sec:psf}
\begin{figure}[!ht]
  \begin{center}
    \epsfig{file=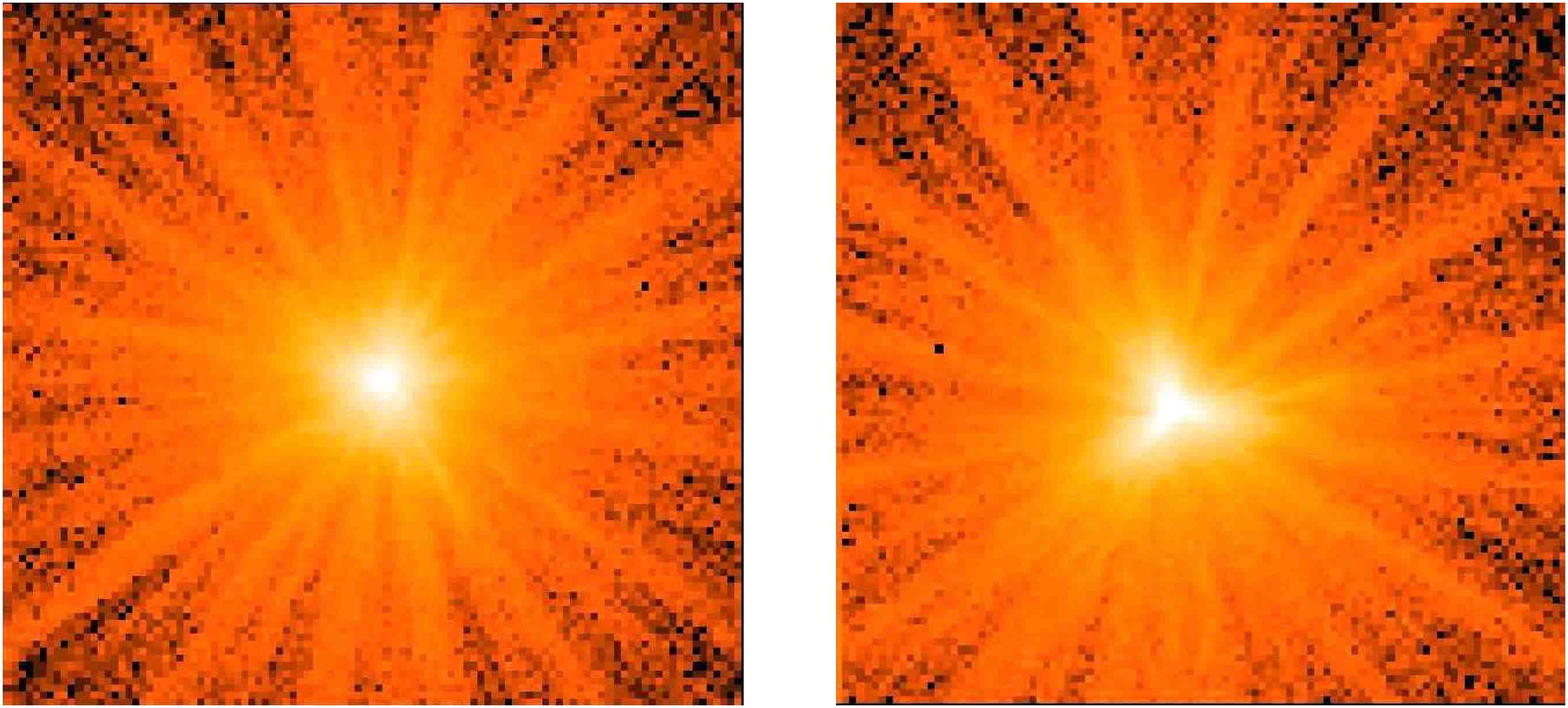, width=8.6cm}
  \end{center}
\caption{The PSF of MOS1 (left) and MOS2 (right) cameras. The images 
are 110\arcsec wide.} 
\label{pferrando-WA2:psf_images}

\begin{center}
    \epsfig{file=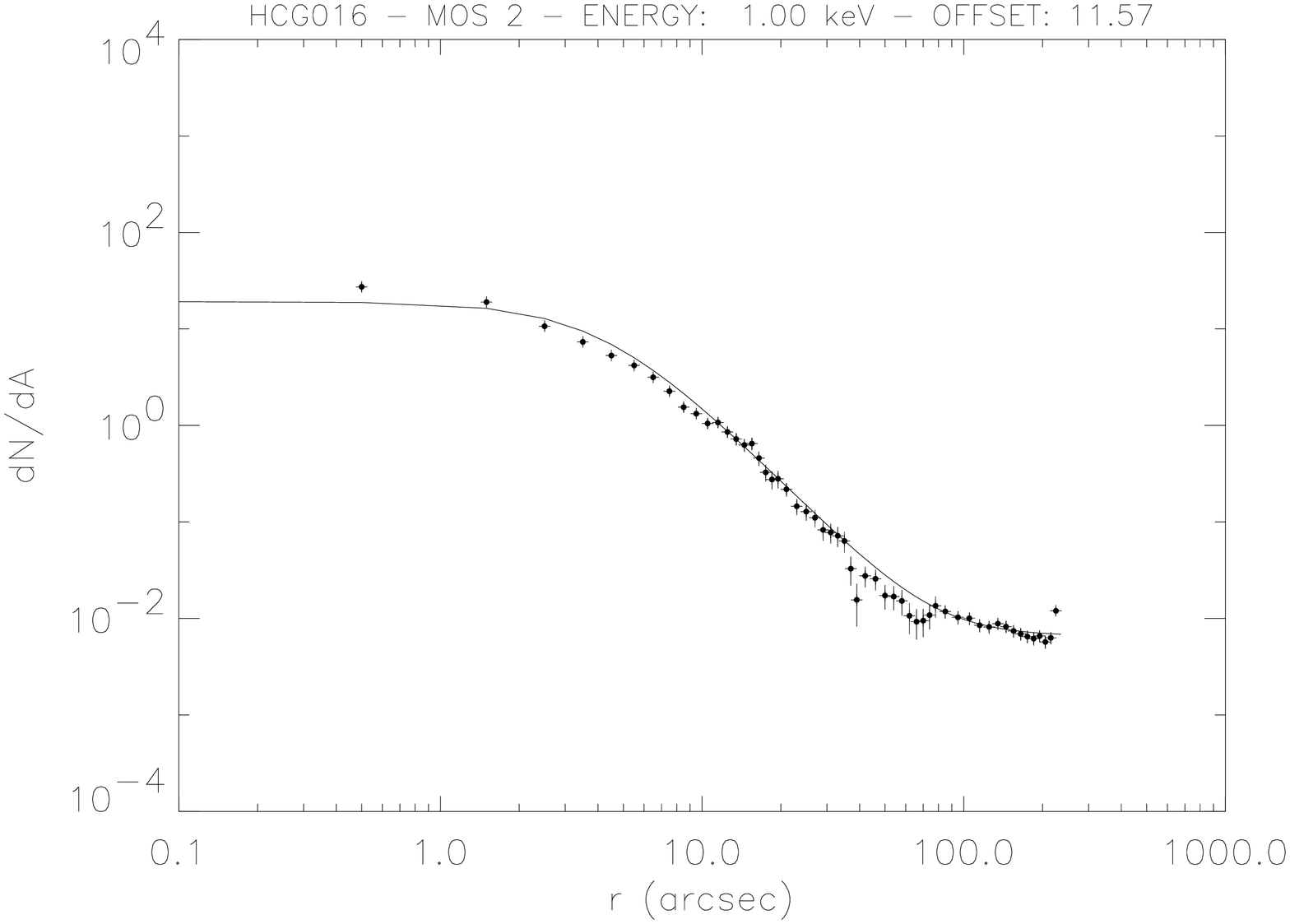, width=6.3cm, angle=90}
  \end{center}
\caption{Example of a PSF fit, at 1 keV for a source being off-axis 
by 12\arcmin}
\label{pferrando-WA2:psf_fit}

\begin{center}
    \epsfig{file=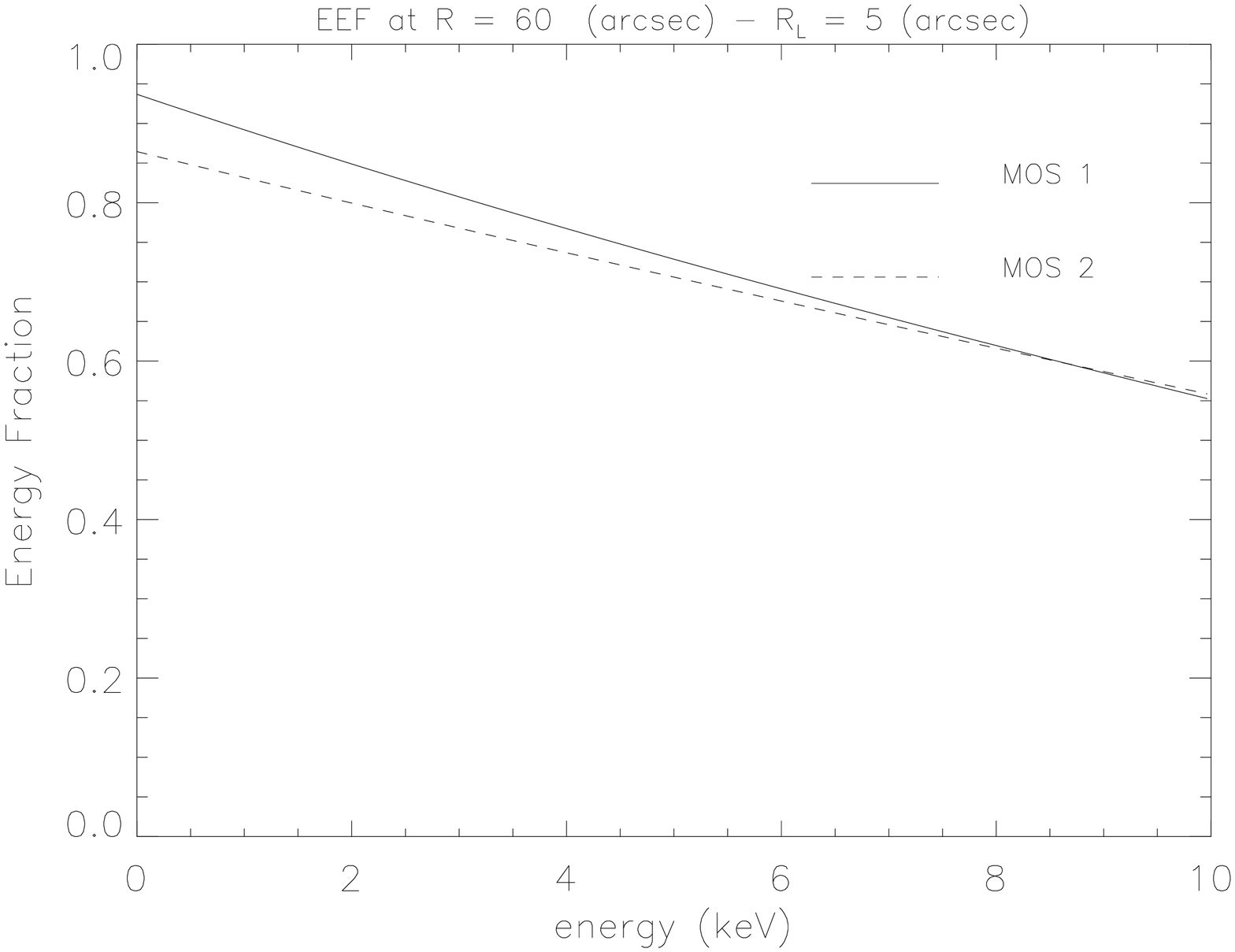, width=6.5cm, angle=90}
  \end{center}
\caption{Encircled Energy Fractions, for an on-axis source, in a ring
with limiting radii of 5\arcsec and 60\arcsec.}
\label{pferrando-WA2:psf_eef}

\end{figure}

The knowledge of the image shape (Point Spread Function, PSF) of a point-like 
celestial source, and of the related Encircled Energy Fraction (EEF)
quantity, are obviously of prime importance regarding the aspects of 
source detection and spectral extraction. The PSFs of the XMM-Newton 
mirrors have a rather complex shape, even for a source located on 
axis, as shown in Figure~\ref{pferrando-WA2:psf_images}. Taking into 
account this complex shape is however unnecessary for most of the 
applications, in which the observer is simply interested in knowing 
the EEF for a given circular extraction radius, or in comparing the 
observed radial profile to the expected one for a point-like source.

For these purposes, \cite*{pferrando-WA2:Ghiz&Mol2002} have looked for 
a simple analytical representation of the PSF. They showed that a 
simple king profile, with only two parameters, represents the data 
very well. An example of fit is given in 
Figure~\ref{pferrando-WA2:psf_fit}. This study was done on a large 
set of data, up to 12 arcmin off-axis position. For each selected 
source, the core radius and the slope parameter were determined for 
seven energy intervals, when allowed by the statistics. 
\cite*{pferrando-WA2:Ghiz&Mol2002} found that these parameters have 
a simple linear dependence on the energy $E$ and off-axis angle 
$\theta$ of the incoming photons. They also provide a range of 
validity, both in $E$ and $\theta$, for this modeling,  based on 
the availability of adequate point source observations. It must be 
noted in particular that the high energy and high off-axis angle 
regions are excluded from the range of application of this model. 
This parametrisation has been fully implemented in the CCFs.

Although the precise core shapes look quite different for the two MOS cameras
(Fig.~\ref{pferrando-WA2:psf_images}), their encircled energy, at 
least on axis, is not drastically different. For example, 
Figure~\ref{pferrando-WA2:psf_eef} shows the on-axis EEF for a ring with 
inner and outer radii of 5 and 60\arcsec respectively, typical of 
extraction regions for piled-up sources. The difference in EEF for the two 
cameras, if not considered properly, would translate into a 
difference of about $\sim 0.03$ on the slope of a power law spectrum.

\subsection{Vignetting}
\label{pferrando-WA2_sec:vignetting}
Compared to the PN camera, the MOS telescopes vignetting is 
complicated by the presence of the gratings arrays. They introduced an 
azimuthal variation, as well as an additional energy dependence above 
about $\sim$~5~keV (\cite{pferrando-WA2:Erd2000}). The in flight 
measurements were performed mainly with observations of  
\object{G21.5-0.9}, on axis and at 10\arcmin and 12\arcmin 
off-axis positions. Although the variations with energy are not completely
understood and are still under investigation, the uncertainty on the 
vignetting corrections is estimated to be better than $\sim$~5~\% up to 
10\arcmin off-axis, and $\sim$~10~\% at 12\arcmin.

\begin{figure}[!h]

\end{figure}\section{Spectral response}

\subsection{Effective area}
\label{pferrando-WA2_sec:effect-area}
\begin{figure}[!ht]
  \begin{center}
    \epsfig{file=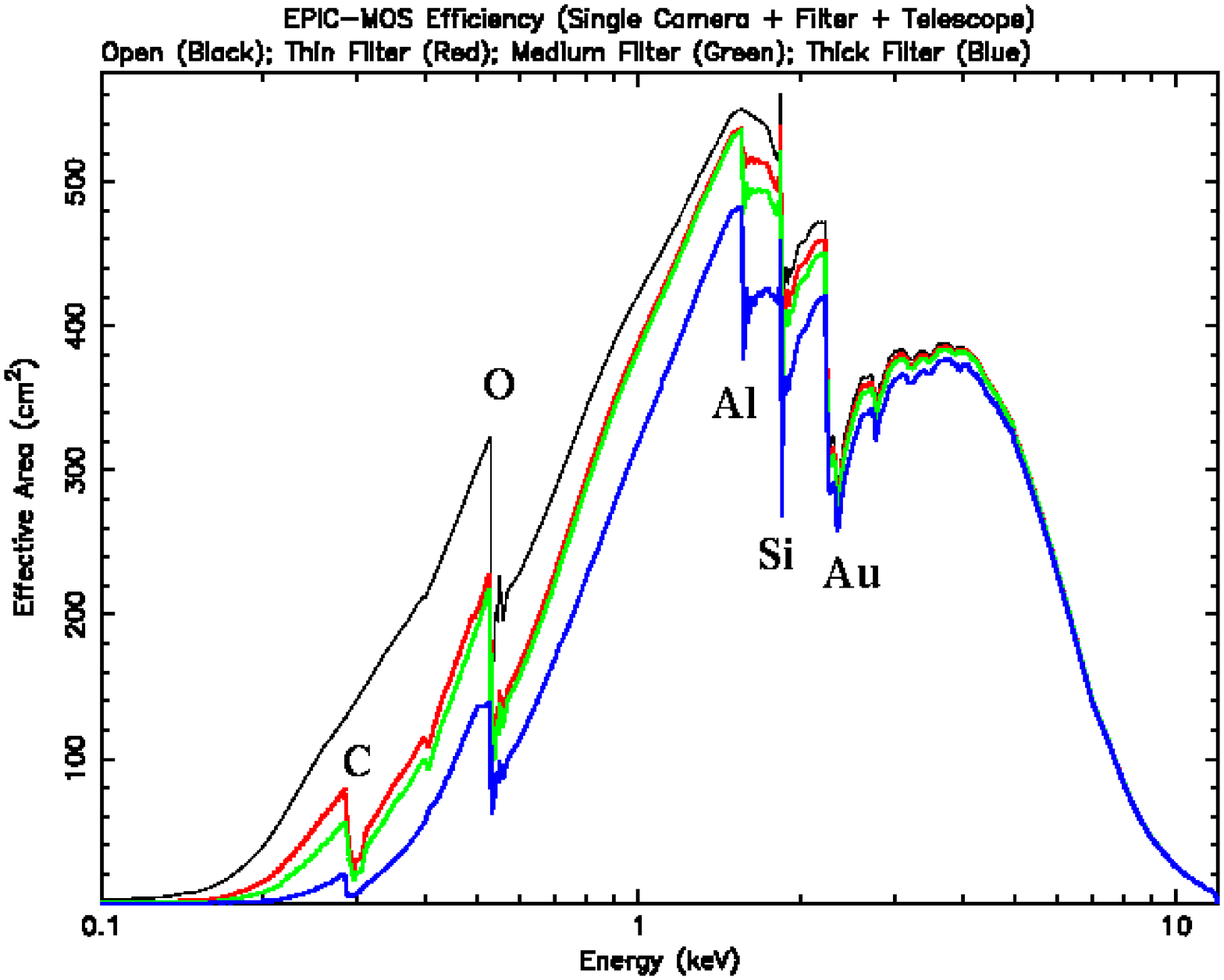, width=8cm}
  \end{center}
\caption{Single EPIC/MOS camera efficiency, for the different 
possible filters. The absorption edges due to the line-of-sight 
material have been indicated.} 
\label{pferrando-WA2:eff_area}
    \begin{center}
    \epsfig{file=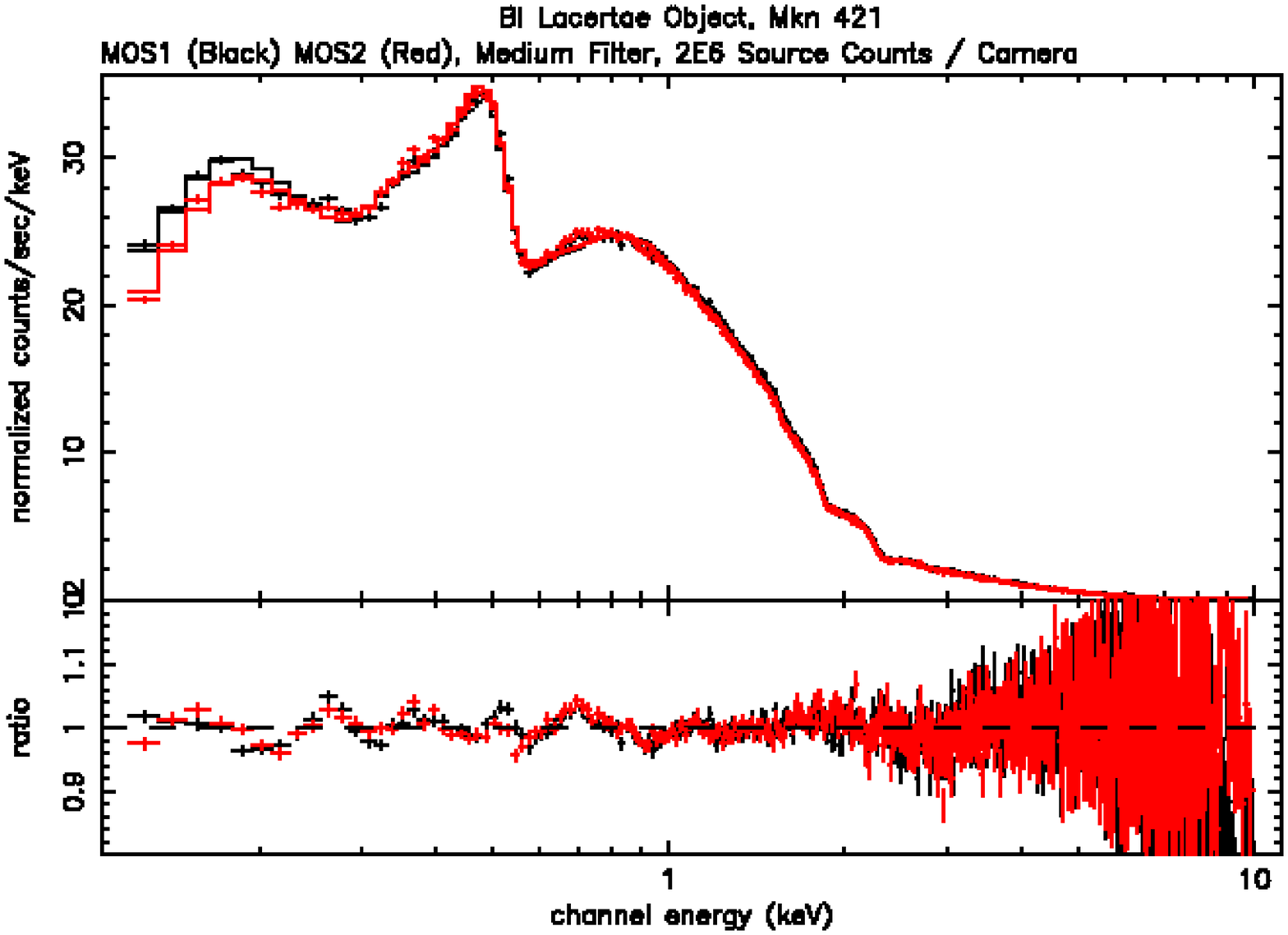, width=8cm}
  \end{center}
\caption{Fit and residuals of Mrk 421 spectrum, for the two MOS cameras.} 
\label{pferrando-WA2:mrkfit}
\end{figure}

The EPIC/MOS effective area, as displayed in 
Figure~\ref{pferrando-WA2:eff_area}, combines the mirror area, the 
gratings and filter transmissions, and the CCD quantum 
efficiency. Starting from ground calibration data for all these quantities, 
\cite*{pferrando-WA2:Sembay2002} have succeeded in working out the corrections 
necessary to minimize the residuals obtained when fitting on-axis sources
with simple spectra, such as BL Lacs. The main changes to the ground 
calibration data made in this work were on the Au edge for the mirror 
response, and the CCD quantum efficiency below one keV, a quantity which 
was poorly determined from the synchrotron measurements. 
Figure~\ref{pferrando-WA2:mrkfit}, from 
\cite*{pferrando-WA2:Sembay2002}, illustrates the quality of the 
fit obtained on a spectrum of \object{Mrk 421}, with more than 
2 millions counts per camera. The residuals are below $5\%$ over
all instrumental edges, and on the full range from $0.2$ to $10$ keV.

Regarding now the overall spectrum, the quality of the fit does obviously 
not ensure the correctness of the parameters. It does however allow 
detailed cross-calibrations between the EPIC instruments to be performed, 
as well as between EPIC and the RGS within XMM-Newton, or between EPIC 
and other X-ray instruments. This is discussed separately in these 
proceedings by \cite*{pferrando-WA2:Griffiths2002}, 
\cite*{pferrando-WA2:Haberl2002}, \cite*{pferrando-WA2:denHerder2002}, and 
\cite*{pferrando-WA2:Snowden2002}. There is generally good agreement 
between the EPIC instruments however small discrepancies still remain, 
namely differences in $n_H$ values (by $\sim 1\times 
10^{20}cm^{-2}$) on the low energy side, and differences of $\sim 
0.1$ for spectral indices on the high energy side. Besides that, 
future work will also be dedicated to the lowest energies, 
below $200$ eV, for which unexpected features have sometimes been
observed.

\subsection{Charge Transfer Inefficiency}
\label{pferrando-WA2_sec:cti}
Although they have been designed to be highly resilient to radiation, 
the MOS CCDs were expected to experience a slow degradation of their 
parallel Charge Transfer Inefficiency due to their constant irradiation by high 
energy protons. By reducing the magnitude of charge packets which 
originates furthest from the read-out node, CTI has the effect of 
shifting energies toward lower values, and to broaden lines. 
The monitoring of the CTI degradation is performed via the measurement 
of Al, Mn-K$\alpha$ and -K$\beta$ lines generated by an internal 
calibration source, which is turned on via a special position of the 
filter wheel (closed w.r.t.\ the sky). These ``calibration source" 
observations have been regularly scheduled at the beginning
of each orbit when XMM-Newton is still in the outer parts of the 
radiation belts in a too high radiation environment to allow 
meaningful astrophysical observations.
\begin{figure}[!ht]
  \begin{center}
    \epsfig{file=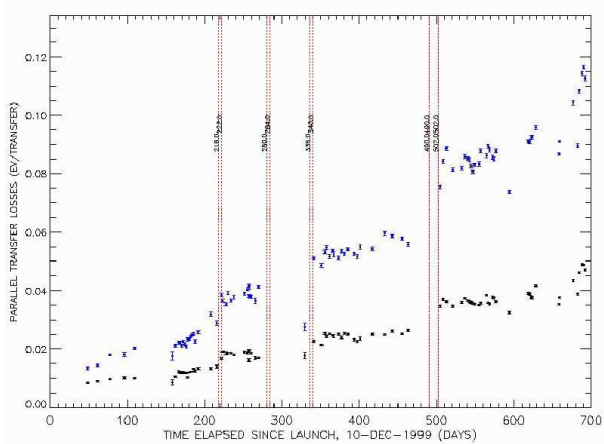, width=8cm}
  \end{center}
\caption{Parallel transfer losses, for the Al and Mn-K$\alpha$ lines 
since launch. The large solar flares times are indicated by the 
vertical lines.} 
\label{pferrando-WA2:cti-evol}
\end{figure}

The detailed study of the CTI evolution is presented by 
\cite*{pferrando-WA2:Bennie2002} in these proceedings. The CTI has been observed to 
increase smoothly with time, with jumps coincident with large solar 
flares, as shown in Figure~\ref{pferrando-WA2:cti-evol}. 
Algorithms have been designed to make a first order correction to the 
CTI effect, and are included in the SAS~5.3 version. Because of its 
statistical nature however, it is not possible to fully recover the 
CTI degradation. \cite*{pferrando-WA2:Bennie2002} have shown that there is still a 
21~\% broadening left on the FWHM of the Mn-K$\alpha$ line measured 
on the full CCD chip, when the SAS~5.3 correction is used. 

Beyond improving the CTI correction in the form currently 
implemented in SAS~5.3, a more drastic improvement under investigation
is to adopt the concept of a deviation map as done for the Chandra ACIS. 
This would allow to account for column to column individualities
which have been evidenced, and reduce the amount of 
unrecoverable damage to $\sim$~5~eV/year for the Mn-K$\alpha$ line 
FWHM, which had a pre-launch value of 125 eV. It remains that the CTI has 
an irreducible line broadening effect, which has now to be taken into 
account in the redistribution function.

\begin{figure}[!ht]
\epsfig{file=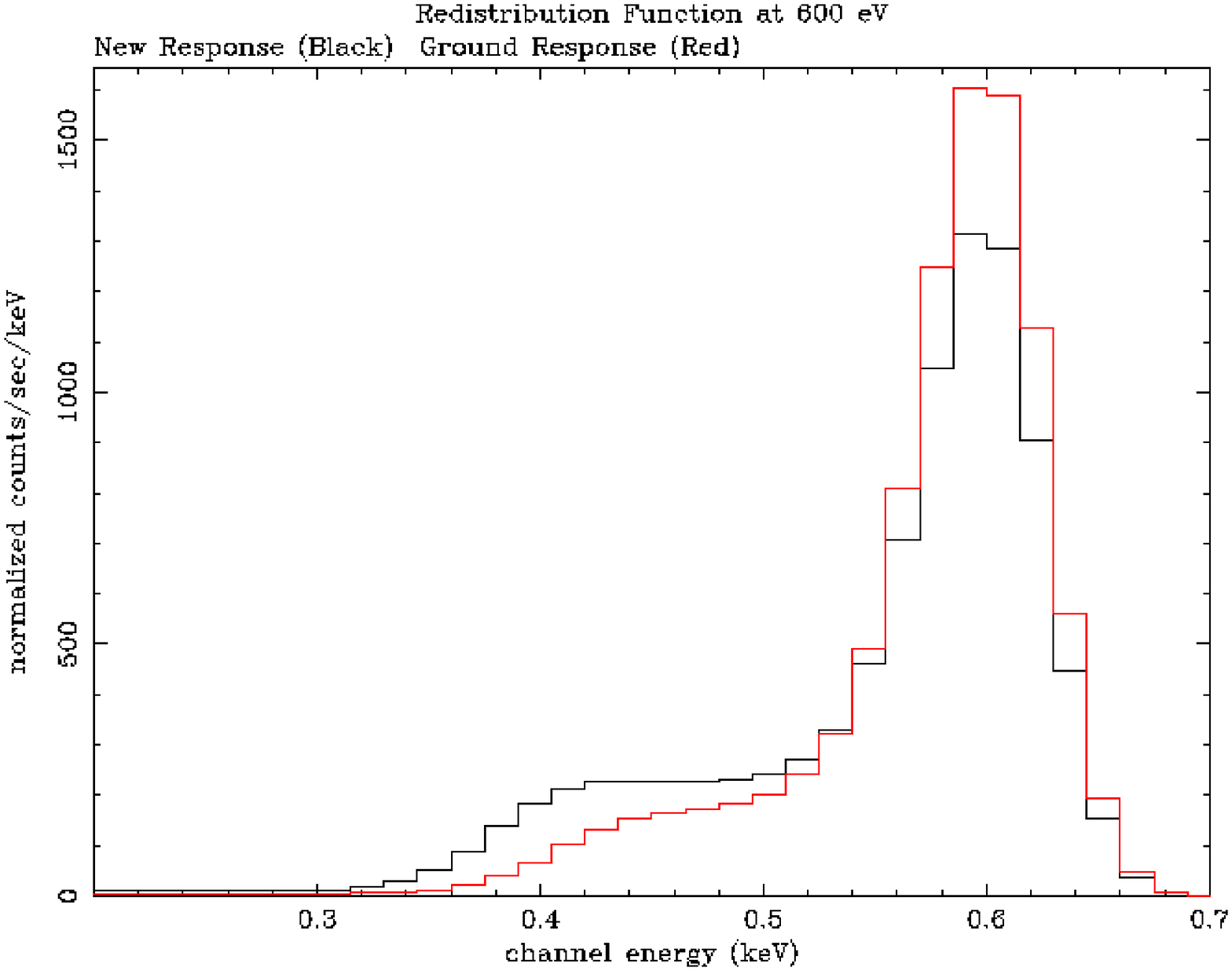, width=4.3cm, height=4.3cm}
\raisebox{4.3cm}{\epsfig{file=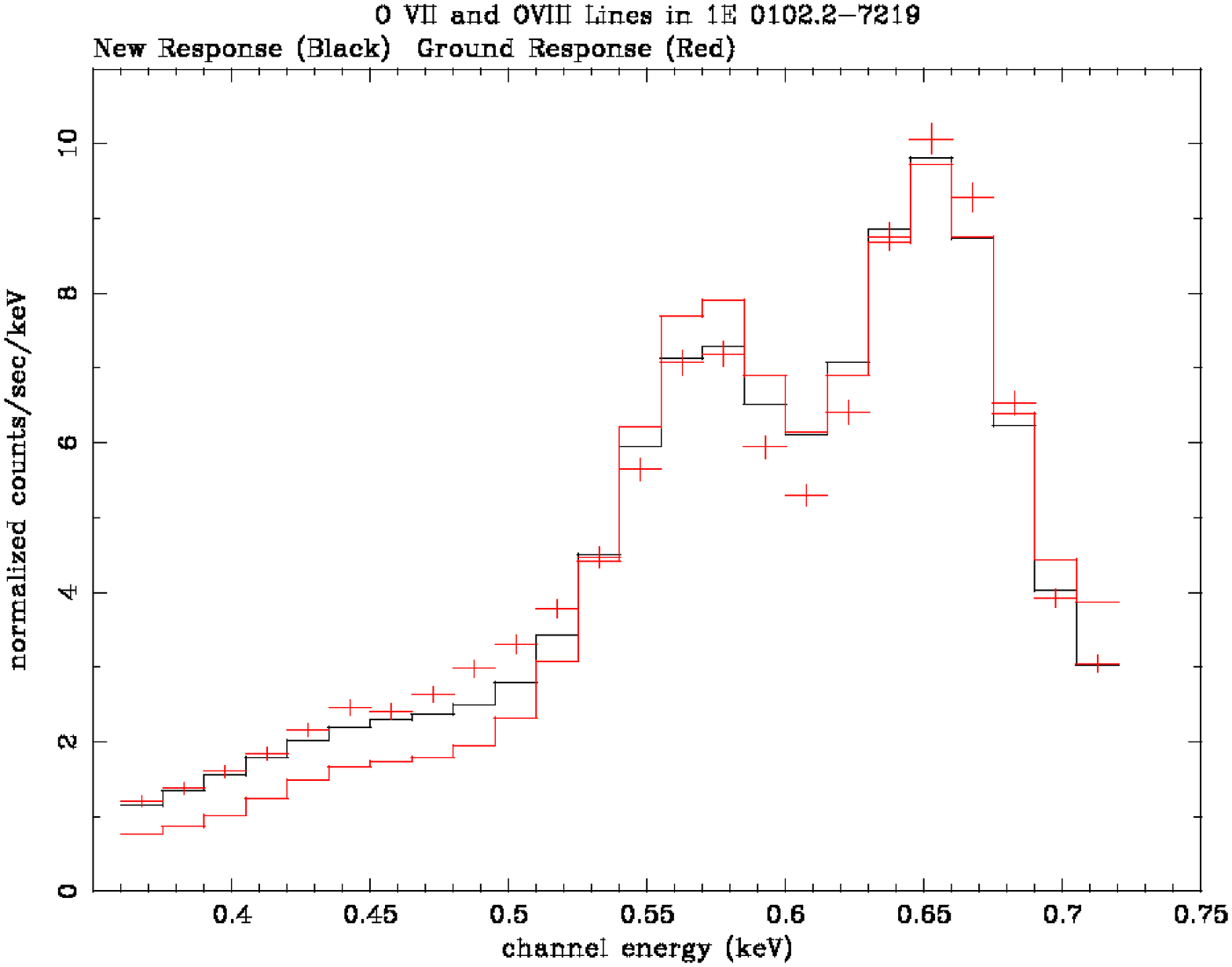, width=4.6cm, 
height=4.3cm,angle=270}}
\caption{Redistribution function at 600 eV (left) and fit to 
\object{SNR~1E0102.2-7219} (right). The red curve corresponds to the ground 
determination of the rmf, the black curve is the shape needed to obtain a 
good fit to the SNR data.} 
\label{pferrando-WA2:rmf.eps}
\end{figure}

\subsection{Redistribution function}
\label{pferrando-WA2_sec:rmf}
The ground calibration, in particular with monochromatic 
beams, has shown that the redistribution function (rmf) is rather complex. 
In particular, below 1 keV it has a significant low-energy shoulder
which becomes increasingly important when the incident energy decreases. 
This is believed to be due to an incomplete charge collection for X-rays 
which are absorbed close to the surface layer 
(\cite{pferrando-WA2:Short1998}).

\cite*{pferrando-WA2:Sembay2002} have been carefully modelling the 
rmf, using sources with strong emission lines, such as 
\object{SNR~1E0102.2-7219}. As is shown in Figure~\ref{pferrando-WA2:rmf.eps}, 
there has been a significant increase in the surface loss component, 
which required modifications to the rmf w.r.t.\ the pre-launch shape. 
The rmf are currently handled together with the effective area has 
stand-alone files. As mentioned above, the CTI degradation implies 
that a time dependent rmf has to be considered now; the way to 
implement that is under investigation. 

Finally, we note here that a single redistribution function 
(as well as effective area) is found to be adequate for describing 
the response of the MOS for all imaging modes; it is not however 
adequate for the timing mode, which spectral response has still to 
be worked out. 

\section{Background}
\label{pferrando-WA2_sec:bkgd}
The background in EPIC has been triggering extensive discussions and 
work inside the EPIC calibration team (and outside\ldots), 
starting right after launch. A lot of work has been devoted to 
characterize this background and to look for ways of reducing it 
to a minimum. The XMM User Handbook gives a pretty complete 
account of the different components of the background found in EPIC, 
MOS and PN, which we will not be repeated here. We simply recall the 
main features of this background, and give the latest developments, in particular those 
from \cite*{pferrando-WA2:Lumb2002a}, \cite*{pferrando-WA2:Lumb2002b}.

It is now recognized that the EPIC background is made of 4 
components, being~:
\begin{enumerate}
  \item A low energy electronic noise,
  \item A transient soft proton ``flare" induced,
  \item A quiet time high energy protons induced,
  \item The diffuse cosmic X-ray background.
\end{enumerate}

The first of this list is occuring at low energy, and is due to 
electronic interferences. It is selectively removed by the standard SAS 
processing. 

The second background is the one causing the strong ``brightening" of 
the image inside the field of view, as in 
Figure~\ref{pferrando-WA2:fig-modes}. It is induced by soft energy 
protons, below a few hundred keV, that directly reach the detectors 
via scattering through the mirrors. It happens in an unpredictable 
way, and has a highly variable light curve when it is present. 
Moreover, it has been shown that its spectral shape is also highly 
variable. Despite numerous attempts, it was impossible to find any 
signature of these events that would allow them to be distinguished, 
and hence filtered out, from real X-rays. In practice, and except 
for very bright sources, periods contaminated by such a background 
have to be excluded from the analysis.

\begin{figure}[!hb]
  \begin{center}
    \epsfig{file=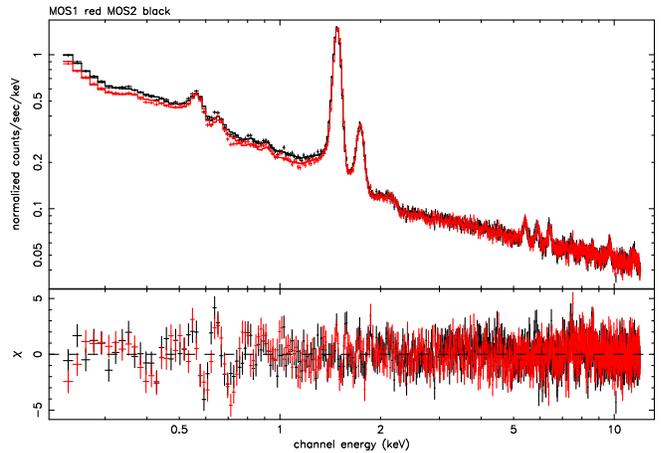, height=8.6cm,angle=270}
  \end{center}
\caption{Spectrum of the internal background, in MOS1 and MOS2} 
\label{pferrando-WA2:bkg-spec}
\end{figure}

\begin{figure}[!hb]
  \begin{center}
    \epsfig{file=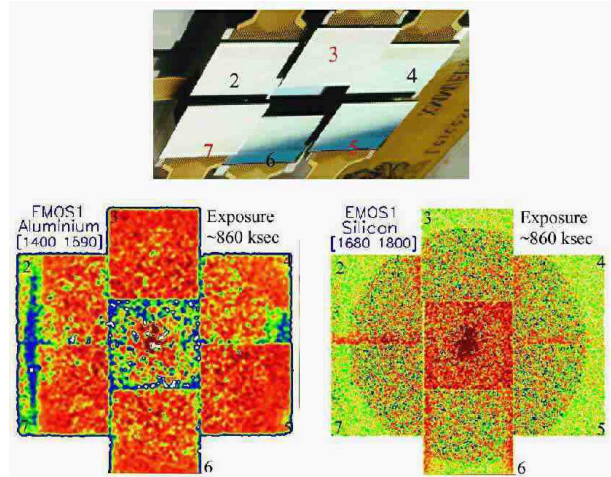, width=8cm}
  \end{center}
\caption{Spatial variations of the Al (left) and Si-K$\alpha$ (right)
lines of the internal background.} 
\label{pferrando-WA2:bkg-lines}
\end{figure}

The characterization of the third background is the one on which 
most of the progress has been done. It was demonstrated to have only 
secular variations, as expected for an origin being from interactions 
of high energy cosmic-rays inside and around the detectors. Its 
spectrum is made of two components~: a continuous component, and a line 
component. The continuous component has a hard spectrum, displayed on 
Figure~\ref{pferrando-WA2:bkg-spec}, probably generated by 
the combination of fluctuations in the energy losses suffered by the 
cosmic-rays when they go through the CCDs, and of Compton electrons 
due to interaction of the gamma-rays produced in the material around 
the CCDs. As expected because of its origin, there is no evidence 
for a spatial variation of this component in the full focal 
plane. The line component is made up 
fluorescent lines, particularly Al and Si. Contrary to the continuous 
component, the distribution of these fluorescent lines, as well as 
weaker ones as Au for example, have a strong spatial dependency. This 
is evidenced in Figure~\ref{pferrando-WA2:bkg-lines} which are maps 
made in the Al and Si lines respectively. They were generated 
from the sum of basically all the PV and CAL phases observations, with 
only SNR observations excluded because of their strong line emissions. 
Because of potential residuals from the on-axis sources, the central 
part of the central CCD should not be considered significant here; 
however all the other structures are significant, and correspond to 
variations of about 15~\% from one area to the other. This is 
readily explained by the localisation of the CCDs, also depicted in 
Figure~\ref{pferrando-WA2:bkg-lines}~: the outer CCDs detect more Al than 
the central one, because of their closer proximity to the Aluminium 
camera housing, while the Si excess seen in the central CCD is 
explained as Si escape lines from the back of the outer CCDs, which 
are located above the central one. The same explanation holds for the 
excess on the edges of CCDs 2, 4 and 6.

In order to cope with this variable background, particularly for 
extended sources, while awaiting for the development of an adequate SAS 
task to model it, \cite*{pferrando-WA2:Lumb2002a} has prepared a 
background template file and has made it available. This data set is 
characterised by i) a long exposure time, over 400~ks, sufficient 
for most observations, ii) an homogeneous collection of observations 
all at high galactic latitudes, with no bright sources, and all with 
the thin filter position. On each of the fields, a source detection 
task was applied and led to about 10 objects per field to be 
excluded, corresponding typically to objects with a flux brighter 
than $1-2\times10^{-14}$ ergs~cm$^{-2}$s$^{-1}$. In the final set 
built by co-adding all observations, the excision of these 
point sources led to only a small $10-15$~\% level of local 
depression. As evidenced in Figures 3 and 4 of 
\cite*{pferrando-WA2:Lumb2002b}, this set gives results very similar 
to the one noted above regarding the fluorescent lines spatial 
dependency.

Finally, by comparing the spectrum derived from this data set 
with a model of the particle induced background spectrum, 
\cite*{pferrando-WA2:Lumb2002b} has found a reasonable power law 
index of 1.4 for the extragalactic background. The same index is also 
found by 
\cite*{pferrando-WA2:deLuca2002}, with a different data set and using a 
somewhat different approach to model the particle background 
modelisation. The absolute normalisation however has still a rather 
large uncertainty.

\section{Areas in progress}
\label{pferrando-WA2_sec:future}
The above sections have stressed the areas for which most of the 
calibration work has been performed. In short, this is basically the 
understanding of the response to on-axis sources, 
observed in all imaging modes, the characterization of the 
telescope vignetting, and of the background components. There are 
a number of calibration areas which have been less touched 
up to now, although this does not mean that they have particular 
problems. Some of them are briefly discussed below.

\subsection{Pile-up}
\label{pferrando-WA2_sec:pile-up}
Although this is not strictly a calibration issue, it is 
important to confirm the conditions for which pile-up has a sizeable 
effect on the spectra. Work is in progress on this area. Preliminary 
results confirm the theoretical analysis of 
\cite{pferrando-WA2:Ballet1999} which stresses that pile-up effects 
are minimized by selection of isolated pixels (pattern 0). Awaiting 
for a potential correction of these pile-up effects, the observer 
has to exclude regions too heavily piled-up for spectral extractions.

\subsection{Off axis analysis}
\label{pferrando-WA2_sec:off-axis}
At high energy, and apart from a slight difference in their depletion 
depth, the EPIC CCDs have homogeneous responses. This is not the 
case below $\sim$ 300 eV, for which ground measurements have shown 
evidence of large inhomogeneities for some of the CCDs. While the 
high energy QE variations from one CCD to the other is implemented in 
the response functions, this is not the case yet of these inhomogeneities 
at very low energy within each CCD. One needs however to verify their 
presence in flight data, a task extremely difficult because of the 
interplay between redistribution and efficiency. Similarly, the thick 
filters have known inhomogeneities below the C edge which are not 
implemented yet.

If a work as extensive as the one performed for on-axis sources has not 
been done yet, and will probably  be impossible to do for the full 
focal plane given the observation time it would request, once can say 
however that the analysis of ``simple" extended sources, as clusters, 
has not revealed any significant problem for the overall off-axis 
response.

\subsection{Timing mode}
\label{pferrando-WA2_sec:timing}
The timing functionalities of the timing mode have been verified to 
be as expected. They are presented by \cite*{pferrando-WA2:Kuster2002} 
in these proceedings. The response function however is quite different 
from that of the imaging modes. This is due to the fact that rows are 
binned within the sequencer program that reads out the CCDs. Also, 
contrary to imaging data, the timing data do not include any information
on the energy in the pixels surrounding the X-ray event. This renders the 
tasks of noise cleaning and energy determination extremely different 
from that of imaging data, and extremely difficult. This is still to 
be done properly, and at the time of writing, the spectra obtained in 
timing mode cannot be accurately modeled.

\section{Conclusions}
\label{pferrando-WA2_sec:conclusions}
We have presented here a short summary of the status of the 
calibration of the MOS cameras of EPIC, as they are less than two years after the 
start of the initial calibration phase. Most of the results have made, or are 
going to make, their way into version 5.3 of SAS. Apart from the 
references given here, the interested reader can find a lot of details 
on the calibration status in the technical notes maintained on the SOC 
web page. 
If the status for on-axis sources can be considered as quite good, 
we have also outlined a number of areas where work is still needed and 
will be done by the calibration team. 

\begin{acknowledgements}
The constant support of PPARC, CEA, CNES and ASI, which fund a very 
large part of the calibration activities on the MOS cameras of EPIC, is 
gratefully acknowledged. 

\end{acknowledgements}

\end{document}